\documentstyle[prd,aps,floats,tabularx,epsfig]{revtex}
\newcommand{\be}{\begin{equation}}
\newcommand{\ee}{\end{equation}}

\newcommand{\bea}{\begin{eqnarray}}
\newcommand{\eea}{\end{eqnarray}}
\newcommand{\bean}{\begin{eqnarray*}}
\newcommand{\eean}{\end{eqnarray*}}

\begin{document}
\draft
\preprint{\
\begin{tabular}{rr}
%CfPA/96-th-15 &  \\
&
\end{tabular}
}
\twocolumn[\hsize\textwidth\columnwidth\hsize\csname@twocolumnfalse\endcsname
\title{Estimate of the Cosmological Bispectrum from the MAXIMA-1\\ 
Cosmic Microwave Background Map}
%\title{An estimate of the Cosmological Bispectrum from the MAXIMA-1 CMB map}
\author{M.~G.~Santos$^{1}$,
  A.~Balbi$^{3,4,5}$,
  J.~Borrill$^{6,4}$,
  P.~G.~Ferreira$^{1}$,
  S.~Hanany$^{7,4}$,
  A.~H.~Jaffe$^{4,8,2}$,
  A.~T.~Lee$^{9,4,5}$
  J.~Magueijo$^{10}$,
  B.~Rabii$^{4,9}$,
  P.~L.~Richards$^{9,4}$,
  G.~F.~Smoot$^{9,4,5,8}$,
  R.~Stompor$^{6,8,2}$,
  C.~D.~Winant$^{4,9}$,
  J.~H.~P.~Wu$^{2}$
}
\address{ $^1$ Astrophysics \& Theoretical Physics, University of Oxford,
Oxford OX1 3RH, UK\\
$^2$
Dept. of Astronomy, 601 Campbell Hall, University of California,
  Berkeley, CA94720-3411, USA
\\$^3$
Dipartimento di Fisica, Universit\`a Tor Vergata, Roma,
Via della Ricerca Scientifica, I-00133,
  Roma, Italy
\\$^4$
Center for Particle Astrophysics, 301 Le Conte Hall, University of
  California, Berkeley, CA94720-7304, USA
\\$^5$
Lawrence Berkeley National Laboratory, 1 Cyclotron Road,
 Berkeley, CA94720, USA
\\$^6$
National Energy Research Scientific Computing Center,
  Lawrence Berkeley National Laboratory, Berkeley, CA94720, USA
\\$^7$
School of Physics and Astronomy, 116 Church St. S.E., University of
  Minnesota, Minneapolis, MN55455, USA
\\$^8$
Space Sciences Laboratory, University of California,
  Berkeley, CA94720, USA
\\$^9$
Dept. of Physics, University of California,
  Berkeley, CA94720-7300, USA
\\$^{10}$ Theoretical Physics, Imperial College, Prince Consort Rd,
London SW7 2BZ, UK}

\maketitle

\begin{abstract}
We use the measurement of the cosmic microwave background taken during
the MAXIMA-1 flight to estimate the bispectrum of cosmological perturbations.
We propose an estimator for the bispectrum that is appropriate
in the flat sky approximation, 
apply it to the MAXIMA-1 data and evaluate errors using
bootstrap methods. We compare the estimated value with what would
be expected if the sky signal were Gaussian and find that it is indeed
consistent, with a $\chi^2$ per degree of freedom of approximately unity.
This measurement
places constraints on models of inflation.
%With this measurement
%we find 
%constraints on deviations from the standard model of inflation and
%on the presence of global topological defects parametrized in terms of
%the O(N) $\sigma$-model.
\end{abstract}
\date{\today}
]
\renewcommand{\thefootnote}{\arabic{footnote}} \setcounter{footnote}{0}
%\section{Introduction}
\noindent
{\it Introduction}:
All theories of structure formation in the universe 
predict the properties of the probability distribution
function (PDF) of cosmological perturbations.
In all cases of interest, the PDF can be completely described
in terms of its spatial n-point correlation functions, which are the expectation
values of all possible products of the random field with itself at
different points in space. Under the assumption of statistical
isotropy and homogeneity,
it is normally more useful to characterize
the PDF in terms of higher order moments of the Fourier
transform of the field. Most readers are familiar with the 2-point
moment, the power spectrum of fluctuations ($C_\ell$). Indeed current
efforts in the analysis of Cosmic Microwave Background (CMB) data
have focused mainly on increasingly precise estimates of the
angular power spectrum. The theoretical
bias for this is clear: for Inflation induced perturbations,
which is the current favourite model of structure formation, the statistics
are Gaussian and all non-zero moments of order $n>2$ can be expressed in
terms of the $C_\ell$. 

In this letter we present the first estimate of the bispectrum
of the CMB on degree, and sub-degree, angular scales.
The bispectrum is the cubic moment of the Fourier transform
of the temperature field and it can be seen as a scale
dependent decomposition of the skewness of the fluctuations (in
much the same way as the $C_\ell$ is a scale dependent decomposition
of the variance of fluctuations). The bispectrum can be used
to look for the presence of a non-Gaussian signal in the CMB sky.
We use the 
data collected with the MAXIMA-1 experiment \cite{hanany} to quantify the 
bispectrum of the CMB.
The Gaussianity of this data set has already been analysed using
complementary methods in \cite{wu}, including the methods of 
moments, cumulants,
the Kolmogorov test, the $\chi^2$ test, and Minkowski functionals
in eigen, real, Wiener-filtered and signal-whitened spaces.

In the past few years, interest in the bispectrum has grown in the
scientific community. Estimates of the bispectrum
in the COBE data proved the statistic to be extremely sensitive to some
non-Gaussian
features in the data, be they cosmological or systematic \cite{heavens};
the quality of galaxy surveys has made it possible to test for the
hypothesis that the matter overdensity is a result of non-linear 
gravitational collapse of Gaussian initial conditions \cite{reds}. 
On the other
hand a serious effort has been undertaken to calculate the expected
bispectrum from various cosmological effects; secondary anisotropies
(such as the Ostriker-Vishniac effect, lensing, Sunyaev-Zel'dovich
effect) \cite{coorayhu}, as well as primordial sources (such as non-linear corrections
to inflationary perturbations or cosmic seeds) may lead to observable
signatures in the bispectrum of the CMB \cite{luo,kw,verde,contaldimag}.

 Let us establish some notation. We shall be working
in the small sky approximation where a map of the CMB can be
considered approximately flat \cite{whiteetal}. 
The anisotropy of the CMB, $\Delta T({\bf x})$,
can then be expanded in terms of 2-dimensional
Fourier modes as follows:
\be
\Delta T({\bf x})=\int\frac{d^2k}{(2\pi)^2}a({\bf k})e^{i{\bf k}\cdot{\bf x}}
\ee
As stated above, the complete statistical properties of $\Delta T$ can
be encoded in the expectation values of products of the $a({\bf k})$.
The power spectrum is defined to be $\langle a({\bf k})a({\bf k'})\rangle=
(2\pi)^2C(k)\delta^2({\bf k}+{\bf k'})$. On small angular scales, the
correspondence between the flat sky power spectrum and the full
sky angular power spectum is straightforward: $C_\ell=C(k)|_{k=\ell}$.
The bispectrum is defined to be
\be
\label{bspdef}
\langle a({\bf k_1})a({\bf k_2})a({\bf k_3})\rangle=(2\pi)^2B({\bf k_1},
{\bf k_2},{\bf k_3})\delta^2({\bf k_1}+{\bf k_2}+{\bf k_3})
\ee
where the delta function constraint is a consequence of the assumption of
statistical
isotropy. 
%In this paper we wish to estimate $B_\ell\equiv B({\bf k_1},
%{\bf k_2},{\bf k_3})$, for the specific configuration
%${k_1}={k_2}={k_3}=\ell$ i.e. highly localized in Fourier space
%(configurations were the ${k_i}$ differ from each
%other are also possible and statistically independent).

%It is $B_\ell\equiv B({\bf k_1},
%{\bf k_2},{\bf k_3})$, for the specific configuration
%${\bf k_1}={\bf k_2}={\bf k_3}=\ell$ , which we wish to estimate in this paper.

\noindent
{\it Method}: 
In this letter we take the approach
adopted by Ferreira, Magueijo \& G\'{o}rski in the analysis of 
the COBE 4 year data \cite{heavens}: we construct
an estimator for the bispectrum, apply it to the MAXIMA-1
data and quantify its variance using Monte Carlo methods.  The
MAXIMA-1 experiment and dataset is described in detail in Ref
\cite{hanany}; as in \cite{wu} we use a map with square
pixels of 8' each.

%We construct the following bispectrum estimator:
Given a map, we  Fast
Fourier Transform it and construct the following bispectrum estimator:
\bea
{\hat B}_{\ell_1\ell_2\ell_3}=&&\frac{1}{N_{\ell_1,\ell_2,\ell_3;\Delta_\ell}}
\sum_{{\bf k_i}\in{\cal S}_{(\ell_i,\Delta\ell)}}
{\cal R}e\left[a({\bf k_1})a({\bf k_2})a({\bf k_3})\right]  \label{eq:estdef}
\eea
with 
\be
{\bf k}_1+{\bf k}_2+{\bf k}_3=0 \label{eq:si}
\ee
where ${\cal S}_{(\ell_i,\Delta\ell)}$ is a ring in Fourier space
centered at $\bf k=0$ and
 with radial coordinates $k\in[\ell_i-\Delta_\ell/2,
\ell_i+\Delta_\ell/2]$, $N_{\ell_1,\ell_2,\ell_3;\Delta_\ell}$ 
are the number of
modes which satisfy this condition
% $R_{[\theta]}$ is a 2-D rotation with an angle $\theta$ 
and ${\cal R}e\{A\}$ is the real part
of $A$. For a given choice of $\ell_i$ (with $i=1,2,3$)
 we obtain an
estimate of the bispectrum
averaged in a bin of width $\Delta_\ell$. 
We correct for the finite resolution of the experiment and 
the pixelization of the map by replacing the quantity
$a({\bf k})$ (that is estimated directly from the map)
by $a({\bf k})/[B({\bf k})W({\bf k})]$, where $B({\bf k})$ and
$W({\bf k})$ are the beam and pixel window functions, 
respectively (see \cite{beam}
for a detailed Fourier space description of the beam).
%To be able to impose the condition in Eq. \ref{eq:si}
% it is necessary to interpolate between values of $ a({\bf k})$.
%We have used linear interpolation between the four nearest neighbours
%on the Fourier grid.

%Note that, in practice, we must take into account the fact that
%the experiment has a finite resolution and that we are
%considering a pixelized map; to correct for these effects
%we replace the $a({\bf k})$ estimated directly from the map
%by $a({\bf k})/[B({\bf k})W({\bf k})]$ where $B({\bf k})$ and
%$W({\bf k})$ are the beam and pixel window functions (see \cite{beam}
%for a detailed Fourier space description of the beam).

There are a number of approximations in our analysis.  We do not
discuss any systematic effects that may have come into play when generating
the map; a detailed description of these effects is presented in
\cite{stompor}.
The flat sky approximation in the estimate of the power 
spectrum is valid to within $1\%$ for the MAXIMA-1 100-square-degrees map.
 The fact that
we are not considering a full sky map leads to two further
complications \cite{hobson}. Firstly, there will be a finite correlation length
in Fourier space between adjacent modes. In Maximum Likelihood
Methods this is automatically taken into account when constructing
the correlation matrix, but in our case we must
take care in assessing how our results depend on the width
of the bins, $\Delta_\ell$, in which we estimate our bispectrum.
Secondly, the map we are working with does not have periodic boundary
conditions, an essential underlying assumption when performing
a Fast  (or Discrete) Fourier Transform. We correct for this by multiplying the
map by a Welch window function which suppresses the mismatch
at the border of the map thus reducing the leakage between
neighboring scales in Fourier space. Naturally we take this
into account when estimating the bispectrum. Finally, the
map to which we apply our estimator will contain
anisotropic instrumental noise, and one may be concerned that
this may bias the estimate. However, given that the
signal and noise are uncorrelated, and the noise is Gaussian
\cite{stompor},
it will not affect our estimate of the bispectrum, merely
its variance. 
%In the next section we will discuss a number of tests
%to check these approximations.
% ============== EDITING TO HERE ================

\begin{figure}
\vskip -.5in
\centerline{\epsfig{file=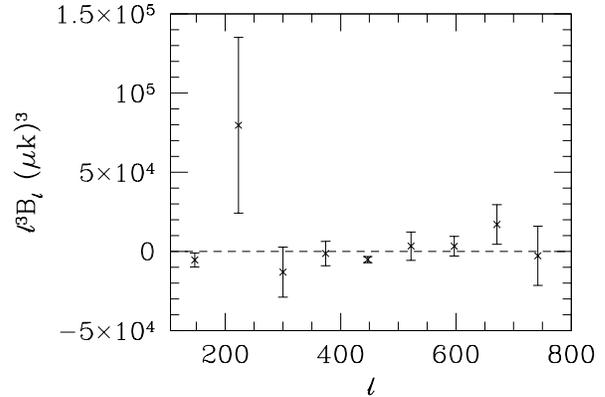,width=3.2in}}
\vskip -.2in
\caption{Estimate of the bispectrum of the MAXIMA-1 CMB map. The
error bars are evaluated using a Monte Carlo bootstrap method. Note that
given the small number of samples for the low $\ell$ components,
there is a large uncertainty in the estimation of the error bars.}
\label{fig1}
\vskip -.2in
\end{figure}

Our goal in this letter is twofold. Firstly to obtain an estimate
of the bispectrum from the data without making any assumptions
about the statistics of the signal and secondly to assess how 
compatible our estimate of the
bispectrum is with the assumption that the MAXIMA-1 data set
is Gaussian.
To obtain a model independent estimate of
the bispectrum and its variance we use bootstrap methods \cite{bootstrap}.
Bootstrap methods are widely used
in situations where one wishes to extract the statistical properties
of a given estimator without making any assumptions about the
distribution from which a sample has been drawn.

One can redefine
the estimator in Eq. (\ref{eq:estdef}) in the following way: divide the
ring in Fourier space into six equally sized angular segments
of width $2\pi/6$; subdivide each of these segments into
$M=2\pi \ell/(6\Delta_\ell)$ angular slices. Within each of these
slices apply the estimator in Eq. (\ref{eq:estdef}), replacing
${\cal S}_{(\ell,\Delta\ell)}$ by the corresponding set of points  within
the slice. Note that this 
is only applicable to the diagonal components of the
bispectrum $B_{\ell\ell\ell}$ (the inclusion of
non-diagonal components will introduce correlations between
samples which will bias bootstrap estimates). In this way
we find $M$ approximately independent estimates of the 
$B_{\ell\ell\ell}$; note that
$\Delta_\ell> 2\pi/({\it field ~size})$ for this to be possible.
If we find the average of these $M$ estimates we recover the value
one obtains by applying (\ref{eq:estdef}). The fact that we have
$M$ (almost) independent estimates puts us in the condition where
one can apply bootstrap methods to estimate the distribution and
consequently the variance. We should note however that there
are two limitations to this approximation.  On the one hand the
sky signal is not uniformly distributed in Fourier
space, i.e. there may be weak correlations between different Fourier modes.
%
%(partial sky coverage will introduce correlations between the a({\bf k}) from
%neighbouring estimates of around $18%$).
%
On the other hand the noise is anisotropic and correlated which
means that the noise covariance matrix is not diagonal
in Fourier space. Both of these effects may lead to correlations
between the M approximately independent samples 
but for large enough $\Delta_\ell$ they should be negligible.
Given that the bootstrap method is the only non-parametric 
(or model independent)
method which one can apply in this situation, we choose to neglect
these correlations \cite{bootstrap}.

Our approach to test for the Gaussianity is
to generate $10^5$ Monte Carlo realizations of the MAXIMA-1 data set, assuming
a Gaussian signal with the power spectrum of the best fit model
to the band powers estimated in \cite{hanany}.
Note that each of these mock data sets will have a realization of the
noise which obeys the full anisotropic, non-diagonal correlation
matrix; moreover the effect of pixelization and finite beam are 
taken into account. We then compare our estimate of the real data
with the Gaussian ensemble and quantify a goodness of fit.

\noindent
{\it Results}: We present the results we have obtained analysing
a square patch in the center of the MAXIMA-1 map, with $50^2$ pixels.
Given the dimensions of the map, we consider $\Delta_\ell=75$; these
correspond to the bin-widths of the estimates of the $C_\ell$ in \cite{hanany}
and lead to correlations of order a few percent between adjacent bins.
In Fig. \ref{fig1} we present the diagonal elements
($\ell_1=\ell_2=\ell_3=\ell$) of the estimate of the bispectrum 
(see also Table \ref{table1}).  Note that
all values of the bispectrum are of order $(0.001-0.01)C_\ell^{3/2}$,
and the fact that $B_{\ell\ell\ell}|_{\ell=224}$ is so large is mostly due to
 the fact that
this corresponds to the peak value of $C_\ell$.

The boostrap errors are evaluated from resamplings with replacement of
the approximately independent samples within each ring; the errors correspond
to the $68\%$ confidence regions with these simulated distributions.
 We find that the average bootstrap errors, $\sigma_{bs}$ over
an ensemble of Gaussian maps to be between $4\%$ and $8\%$ lower
than the true underlying variance. This bias is due to
the correlations between adjacent samples within each ring.
Moreover
the number of approximately independent samples ranges from $M=2$ at
$\ell=148$ to $M=10$ at $\ell=748$ and one should therefore bear in
mind that, for low $\ell$ the variance in the estimate of
the bootstrap errors is large.

\begin{figure}
\vskip -.5in
\centerline{\epsfig{file=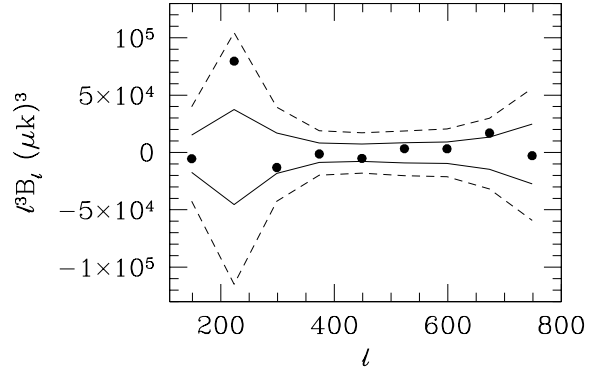,width=3.2in}}
\vskip -.2in
\caption{Comparison with a gaussian sky. The solid (dashed) lines delimit
the 68$\%$ (95$\%$) confidence region determined from a Monte Carlo
simulation of a Gaussian sky; the MAXIMA-1 noise covariance matrix
was used to simulate realistic, anisotropic noise and the beam and
pixel window functions were included.}
\label{fig2}
\vskip -.2in
\end{figure}

We have performed a number of tests to evaluate how robust the result is
on the parameters of our estimator. We have taken a larger patch of
the MAXIMA-1 map and considered maps of $50^2$ pixels with
different locations within the MAXIMA-1 map. The estimated bispectra
vary by a few percent. Alternatively we have considered different
bin-widths (with $\Delta_\ell=60$ and $\Delta_\ell=90$) and found
that estimates of the bispectrum vary smoothly and are consistent within
different binnings. The use of the Welch window function turns out
to be essential for small $\ell$; this is to be expected as it should
be the values of $B_{\ell\ell\ell}$ for low $\ell$ which are most affected
by finite size effects. A different choice of window function (such as
the Bartlett window function) changes the estimate of 
$B_{\ell\ell\ell}|_{\ell=148}$
and $B_{\ell\ell\ell}|_{\ell=224}$ 
by an order of $15\%$ but leaves the remaining values of
$B_{\ell\ell\ell}$ 
unaffected. One final test we have undertaken was
to rotate the ring considered in Fourier space, this way displacing
the $M$ angular slices;
we have found that the results vary by at most 10$\%$ in the lowest
$\ell$ bin.

In Fig \ref{fig2} we plot the diagonal estimate of the MAXIMA-1 bispectrum
compared to the 68$\%$ and 95$\%$  contour
 values if the sky was indeed Gaussian.
We have checked that our statistic is unbiased even in the presence
of anisotropic Gaussian noise and, as can be seen,
the MAXIMA-1 $B_\ell$ seem to be
consistent with the Gaussian assumption.
The obvious way to quantify this is to use a goodness of fit.
For the Monte Carlo realizations of the Gaussian sky signal, we
find that most of the histograms of  the $B_{\ell_1\ell_2\ell_3}$ 
are well approximated by Gaussians
and we therefore define the standard
$\chi^2=\sum_{\ell_1\ell_2\ell_3\ell_1'\ell_2'\ell_3'}(B^{obs}_{\ell_1\ell_2\ell_3}-B^{th}_{\ell_1
\ell_2\ell_3}){\cal C}^{-1}_{\ell_1\ell_2\ell_3\ell_1'\ell_2'\ell_3'}
(B^{obs}_{\ell_1'\ell_2'\ell_3'}-B^{th}_{\ell_1'
\ell_2'\ell_3'})$ where $B^{th}_{\ell_1\ell_2\ell_3}=0$ 
and ${\cal C}$ is the covariance
matrix of the estimators evaluated from the Monte-Carlo realizations.
In all we have 115 values and we find 
$\chi^2=130$. From $10^4$ realizations we construct the
expected distribution of this $\chi^2$:
we find that $70\%$ of the
distribution is contained to the left of the measured value. 
Even if we remove the outlier from the set of bins centered
at $\ell=224$ we still find that $52\%$ of the distribution
lies to the left of the measured $\chi^2$.

\noindent
{\it Cosmological Implications:}
%Let us now assess the cosmological consequences of our constraint.
One
can roughly divide the two possible sources of non-gaussianity in
the CMB into primordial and late time. The latter have been extensively
studied in \cite{coorayhu} and typically give rise to non-zero
bispectra on very small angular scales ($\ell>1000$). We do not expect
to find any evidence for such signatures in the MAXIMA-1 map.
Moreover, the observed
bispectrum limits point source contribution to the MAXIMA power spectrum as it
shows no significant rise at high $\ell$.
Primordial effects may give rise to non-Gaussianity on degree
scales and we shall focus on a few possibilities now. Inflation
predicts almost Gaussian fluctuations to a very good approximation;
there is however the possibility that second order corrections in
the evolution of the inflaton field may lead to mild non-Gaussianity.
Komatsu and Spergel \cite{kw} have parameterized this non-linearity
in terms of a ``non-linear coupling constant'', $f_{NL}$, which can
be related to dynamical parameters in a variety of models of
inflation. For example, $f_{NL}\simeq (3\epsilon-2\eta)$ where
$\epsilon$ and $\eta$ are the slow roll parameters of single field
inflation; one expects from slow roll models that at most
$f_{NL}\simeq {\mathcal O}(1)$.
An order of magnitude estimate gives 
\begin{eqnarray}
B_{\ell_1\ell_2\ell_3}&\simeq& b_{\ell_1\ell_2}+
b_{\ell_2\ell_3}+b_{\ell_3\ell_1} \nonumber \\
b_{\ell_i\ell_j}&=&-\frac{1.1\times 10^{2}}{T_{CMB}}
\,f_{NL}\,\left(\frac{{\Delta T}_{\ell_i}}{\ell_i}\right)^2
\left(\frac{{\Delta T}_{\ell_j}}{\ell_j}\right)^2 \mbox{.}
\nonumber
\end{eqnarray}
and ${\Delta T}_{\ell}^2\equiv\ell\,(\ell+1)\,C_\ell/(2\pi)$. Using the Monte
Carlo realizations described before it is possible to estimate the smallest
amplitude, $|f_{NL}|$, distinguishable from the Gaussian hypothesis; we find
that $|f_{NL}|<944$ is indistinguishable from a Gaussian signal at the $95\%$
confidence level.
Note that the use of lower multipoles (as measured by COBE)
should narrow this interval.
A fit to the measured values using the Gaussian covariance matrix
 gives
$|f_{NL}|\simeq 900$.
($\chi^2=122$).
%The number of independent estimates contributing to the bispectrum
%for $\ell=148$ is quite low and if we don't consider this point the fit gives
%$f_{NL}\simeq 820 \pm 2070$.
% with $\chi^2=14.7$.

%\begin{figure}
%\vskip -.5in
%\centerline{\epsfig{file=fig_bi3.ps,width=3.2in}}
%\vskip -.2in
%\caption{The goodness of fit of the measured $B_\ell$ as compared
%to the expected distribution for $10^5$ Gaussian skies. We find that 32\%
%of the curve is to the left of the measured value of $\chi^2=6.6$
%(indicated by a vertical dashed line). }
%\label{fig3}
%\vskip -.2in
%\end{figure}

More exotic possibilities can be considered, such as for example, global topological defects.
 A semi-analytic framework exists which allows one to
calculate the statistical effects using the $O(N)$ non-linear $\sigma$-model.
Different values of $N$ will correspond to different types of localized objects,
with, for example, $N=2$ corresponding to global strings, $N=3$ monopoles and $N=4$ 
corresponding to textures (taking $N$ to infinity we recover gaussianity).
 Verde {\it et al}
\cite{verde} (see also \cite{luo}) have estimated 
the bispectrum and found that
\begin{eqnarray}
B_{\ell_1\ell_2\ell_3}
\simeq \frac{2.0\times 10^{5}}{T_{CMB}}\,\alpha\,\left(
\frac{{\Delta T}_{\ell_1}}{\ell_1}\frac{{\Delta T}_{\ell_2}}{\ell_2}
\frac{{\Delta T}_{\ell_3}}{\ell_3}\right)^{\frac{4}{3}}
\mbox{,}\nonumber
\end{eqnarray}
 where $\alpha=N^{-1/2}$ (we should point out that
this expression was derived for large angles). In what follows
we shall extrapolate this expression to subdegree scales. 
% Applying this to the gaussian estimates
%as above gives $\alpha <2.17$ so that $N \ge 1$.
The current sensitivity is such that models with
$\alpha \simeq < 2.4$ are indistinguishable from Gaussian
theories; this range of $\alpha$ corresponds to any value 
of $N$. We find the best fit $\alpha$ to be $\alpha=2.2$.
Current estimates of the bispectrum do not therefore constrain
global topological defects.

The bispectrum analysis of the MAXIMA data indicates that the data
is consistent with Gaussianity. This reinforces the conclusions obtained
in \cite{wu} and validates the assumptions that go in to the data-analysis
pipeline, namely the assumption of Gaussianity of the sky signal which
goes into both Maximum-Likelihood and Monte-Carlo estimates of the
power spectra. 

\begin{table}
\begin{center}
\begin{tabular}{|c|c|c|c|c|}
\hline $[\ell_{min},\ell_{max}]$ & $\ell^3B_{\ell} $ &
 $\ell^3\sigma_{bs} $ & $\ell^3\sigma_{G} $ & $\ell^3\sigma_{N} $\\
\hline 	$[111,185]$ & -5455 & 4477 & 16329 & 38\\
	$[186,260]$ & 79622 & 55440 & 41363 & 145\\
	$[261,335]$ & -13167 & 15798 & 17590 & 183\\
	$[336,410]$ & -1373 & 7687 & 8504 & 366\\
	$[411,485]$ & -5208 & 1977 & 7593 & 1071\\
	$[486,560]$ & 3298 & 8939 & 8801 & 1815\\
	$[561,635]$ & 3199 & 6213 & 9387 & 2892\\
	$[636,710]$ & 16952 & 12518 & 13997 & 5939\\
	$[711,785]$ & -2802 & 18725 & 26058 & 14197\\

\hline
\end{tabular}
\end{center}
\caption{Measured bispectrum values and corresponding errors.
The first column has the bandwidths, the second column the estimate
of the bispectrum, the third column has an estimate of its
variance using bootstrap methods, the fourth
column has an estimate of its variance assuming the signal is Gaussian
and the fifth column
has its variance just due to noise. Columns 2-5 are in units of $(\mu$K$)^3$.}
\label{table1}
\vskip -.2in
\end{table}
\vskip .3in
{\it Acknowledgments}: MGS acknowledges support from FCT (Portugal).
%Funda\c{c}\~ao
%para a Ci\^encia e a Tecnologia.
JHPW and AHJ acknowledge support from
NASA LTSA Grant no.\ NAG5-6552 and NSF KDI Grant no.\ 9872979.
PGF acknowledges support from the Royal Society.
RS and SH acknowledge support from
NASA Grant NAG5-3941.
BR and CDW acknowledge support from NASA GSRP
Grants no.\ S00-GSRP-032 and S00-GSRP-031.
MAXIMA is supported by NASA Grant
NAG5-4454 and by the NSF through the CfPA
%Center for Particle Astrophysics 
at UC Berkeley, NSF cooperative agreement AST-9120005.

\tighten

\end{document}